\documentclass[amsmath,amssymb,amsfonts,nofootinbib]{revtex4}

\usepackage[colorlinks]{hyperref}
\usepackage[nameinlink]{cleveref}
\Crefname{figure}{Fig.}{Figs.}

\usepackage[paperwidth=210mm,paperheight=297mm,centering,hmargin=2.0cm,vmargin=1.0cm]{geometry}
\usepackage{xcolor}
\usepackage{graphicx}
\graphicspath{{images/}}
\usepackage{subfig}
\usepackage{float}
\usepackage{placeins}
\newcommand{\RomanNumeralCaps}[1]
    {\MakeUppercase{\romannumeral #1}}
    
\usepackage{datetime}
    
\begin{document}
\title{T-odd anomalous interactions of the top-quark at the Large Hadron 
Collider}
\author{Apurba Tiwari and Sudhir Kumar Gupta}   
\email{atiwari@myamu.ac.in, sudhir.ph@amu.ac.in}

\affiliation{Department of Physics, Aligarh Muslim University, Aligarh, UP -- 
$202 002$, INDIA}  

\begin{abstract}
We study the effects of T-odd interactions of top-quark via the pair production 
of top-quark in the semileptonic detection modes 
at the Large Hadron Collider by means of the T-odd observables 
constructed through the momenta of the observed decay products of the 
top (and anti-top)-quark for a wide range of CP-violating scale $\Lambda$. Estimates on sensitivities of the coupling strength of 
such interactions for 13 TeV LHC energy with $\int \mathcal L dt$ = 36.1 fb$^{-1}$, 140 fb$^{-1}$ and for HL-LHC with 14 TeV energy with integrated luminosities of 0.3 ab$^{-1}$, 1 ab$^{-1}$, 2 ab$^{-1}$ and 3 ab$^{-1}$ are also presented for
 $\Lambda$ ranging between $M_W$ and 2 TeV.  
\end{abstract}

 \date{\today ~~\currenttime}
 
  \maketitle

\section{Introduction} \label{intro}

The phenomenon of charge and parity violation which was originally discovered in the neutral 
kaon-system via measuring the oscillation probability of $K^0$ into 
${\bar K}^0$ \cite{Christenson:1964fg} is now well understood. It besides 
being a new effect, had provided the ground for further exploration not only 
as an independent phenomenon but also its relation with a phenomenon such 
as Leptogenesis 
\cite{Chun:2017spz,Antusch:2017pkq,Achelashvili:2017sro,Moffat:2018smo}, 
Baryogenesis \cite{Cui:2014twa}, nature of the Higgs boson 
\cite{Barbiellini:1979ja,Vainshtein:1980ea} and Dark matter of the Universe \cite{Raffelt:1997de,Halzen:1991kh,Kamionkowski:1998is}. The Standard-Model (SM) which is originally $CP$-symmetric could still 
allow a tiny amount of $CP$-violation via the inter-generational mixing of 
the fermions having identical quantum numbers through CKM-matrices 
\cite{Krawczyk:1987wm}. However such effects are not 
sufficient to provide a satisfactory explanation to the observations such 
as the finite though a tiny amount of electric-dipole-moment of the 
neutron \cite{Bhattacharya:2018qat, Mereghetti:2018oxv}, origin to which 
may lie in the violation of $CP$-symmetry in the strong sector. These, therefore, 
require one to explore the possible sources of $CP$-violation beyond the 
Standard Model.

Guided with the aforementioned phenomenon, in the present article we 
explore the possibility of a model-independent extension of the SM in 
the form of T-odd anomalous interactions of the top-quark with gluons in 
the context of top-pair production at the LHC with pre-existing data at 
13 TeV center-of-mass (C.M.) energy and the forthcoming 14 TeV run for 
projected Luminosities of about 0.3 ab$^{-1}$, 1 ab$^{-1}$, 2 ab$^{-1}$ and 3 ab$^{-1}$ respectively.

The T-violating interactions of the top-quark have already been studied in the literature for a fixed $CP$-violating scale in the 
Refs.~\cite{Bernreuther:2017cyi,Hagiwara:2017ban,Aaboud:2016bmk,Hayreter:2015ryk, 
Saha:2015lna,Berge:2015naf,ATLAS:2013ula,deVries:2018mgf,Cirigliano:2016nyn,Chien:2015xha,Cheung:1996kc,
Cheung:1995nt,Li:2017esm,Gupta:2009eq,Antipin:2008zx,Khachatryan:2016ngh,Monfared:2016vwr,Dawson:2013owa,Dwivedi:2016xwm,Bernreuther:2010ny}; for example, $CP$-violation at 
future $e^+e^-$ collider in $t\bar{t}$ production is 
investigated in Ref. \cite{Bernreuther:2017cyi}, Ref. \cite{Hagiwara:2017ban} considered $CP$-violation due to complex top-Yukawa 
coupling in $e^+e^- \to ht\bar{t}$ at future $e^+e^-$ 
collider, Charge-asymmetries in $b\bar{b}$ pair from 
top-quark decay were first analysed in Ref. \cite{Aaboud:2016bmk}, Ref. \cite{Hayreter:2015ryk} studied the $CP$-violation using T-odd correlations in lepton plus jets channel, Ref. \cite{Saha:2015lna} explored the possibilities of $CP$-violation in a rare process of top decay $t \to b\bar{b}c$, Ref. \cite{Berge:2015naf} examines 
the possible $CP$-violating effects due to one-loop corrections to the 
top pair production process in the complex MSSM with minimal flavor 
violation (MFV) at hadron colliders and Ref. \cite{ATLAS:2013ula} 
investigates the $CP$-violation in the decay of a single top-quark produced in 
the t-channels. Similar studies have been performed for effective anomalous CP-violating couplings for the process $\gamma \gamma \to t\bar{t}$ in Refs. \cite{Poulose:1998sd,Grzadkowski:2003tf}, at FLC in Ref. \cite{Lietti:2000dz} and in the context of muon colliders in Refs. \cite{Hioki:2007jc}. The present article explores the effect of such anomalous 
interactions for a wide range of $CP$-violating scale and provides the 
LHC-sensitivities for the coupling of such interactions via 
the process $pp \to t\bar{t} \to 
(bl^+\nu_l)(\bar{b}l^-\bar{\nu_l})$ using 
T-odd triple product correlations defined in Ref. \cite{Valencia:2013yr}.

Plan of the article is as follows: In section 
\ref{T-oddint} we discuss the model and 
possible T-odd observables for the top-pair production at the LHC and 
how these observables are suitable for analysing the effects of the $CP$-violation. Section 
\ref{Numanl} discusses the numerical procedure and results on T-odd interactions. The experimental sensitivities of the T-odd couplings are also discussed in the same section. Finally, we summarise our findings in section \ref{summary}.
%%%%%%%%%%%%%%%%%%%%%%%%%%%%%%%%%%%%%%%%%%%%%%%%%%%%%%%%%%%%%%%%%%%%

\section{T-odd observables and top-pair production} \label{T-oddint}

$CP$-violation in the quark sector (except for the top-quark) faces an observational difficulty which partially lies in the fact that due to relatively larger life-time than the hadronisation scale, which is of about 140 MeV ($\simeq m_{\pi^0}$, the mass of pion), quarks form bound states and thereby 
leave no scope for studying pure $CP$-violation. By being much heavier than other quarks and also much energetic than the hadronisation scale, top-quark turns out to be the only expectation to test direct CP-violation in the quark sector. The life-time of a top-quark is less than the time required for a quark to hadronise
therefore it does not form any bound state. Consequently the dynamics of top-production and decay do not get affected by complications of non-perturbative and bound state physics and, therefore, the CP-violation effects involving top-quark will be of direct type. At hadron colliders, processes 
involving top-quarks have a further advantage in having larger 
cross-sections due to the strong interactions. This, therefore, enables us 
to directly investigate the effects of such interactions via the 
pair-production of the top-quarks and their subsequent decays into 
a pair of leptons and b-quarks.

Our study of finding $CP$-violation is based on estimating asymmetries through CP-violating observables. $CP$-odd observables can be formed using T-odd correlations which may not necessarily be CP-odd instead these could be CP-even as well and T-odd is not for time-reversal here, rather, it represents naive T-odd \cite{Han:2009ra}.

The chromo-electric dipole moment (CEDM) of the top-quark causes the CP-violation in the top pair production vertex. In the presence of T-odd interactions of top-quark with gluon, the SM Lagrangian could be modified for $t\bar{t}$ production process by the following interaction term \cite{Gupta:2009wu}

%%%%%%%%%%%%%%%%%%%%%%%%%%%%%%%%%%%%%%%%%%%%%%%%%%%%%%%%%%%%%%%%
\begin{eqnarray}
\label{intlag}
\mathcal L_{int}&=& 
-i\frac{g_s}{2}\left(\frac{d_g}{\Lambda}\right)\bar{t}\sigma_{\mu\nu}\gamma_5\ G^{\mu\nu}\ t, 
\end{eqnarray}
%%%%%%%%%%%%%%%%%%%%%%%%%%%%%%%%%%%%%%%%%%%%%%%%%%%%%%%%%%%%%%%%%

with $g_s$ being the strong coupling constant, $G^{\mu\nu}$ the gluon field-strength tensor, ${d_g}$ and $\Lambda$ being the interaction strength and energy scale of the $CP$-violation respectively and $\sigma_{\mu\nu} = 2i[\gamma_{\mu}, \gamma_{\nu}]$. The Lagrangian in Eq. \ref{intlag} will give a new dimension five vertex $t\bar{t}gg$ (which is absent in the SM) in addition to modifying the pre-existing $t\bar{t}g$ vertex. This new vertex $t\bar{t}gg$ is obviously $CP$-odd in nature according to the above equation.

\begin{figure*}[ht!]
\centering
\includegraphics[width=\textwidth,height=8.5cm]{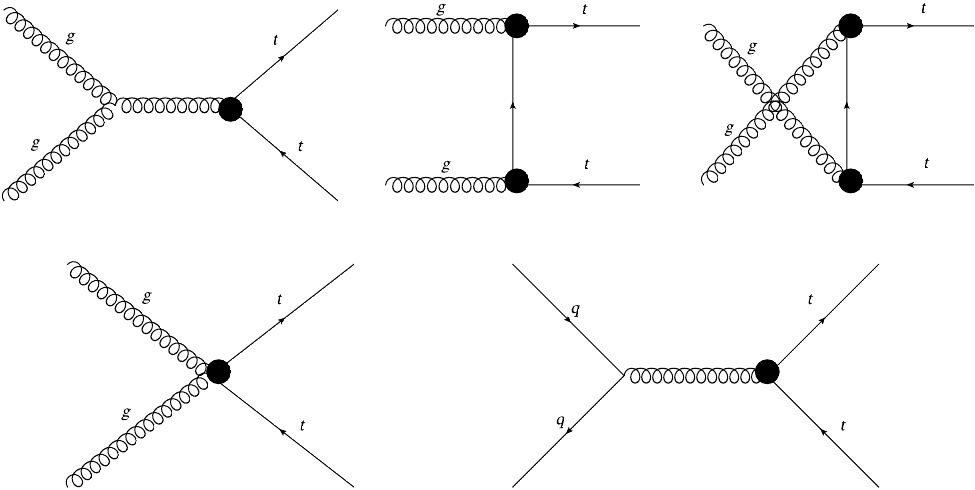}
\caption{\small Feynman diagrams responsible for top-quark pair-production at the LHC.}
\label{topprod}
\end{figure*}

These would clearly have a significant contribution to the top-pair 
production processes at hadron colliders, particularly for colliders 
alike LHC where the fusion of gluons emerging from the colliding protons 
makes about 90$\%$ contribution, the rest being the annihilation of 
light-partons of opposites charges. A schematic representation of 
various parton-level processes describing the production of $t\bar{t}$ at the 
LHC where the modification occurs due to the presence 
of additional T-odd interactions given by Eq.~\ref{intlag} shown in Figure \ref{topprod}. The first four diagrams of Figure \ref{topprod} represent the production of $t\bar{t}$ pairs through $gg$ fusion and the last one is via $q\bar{q}$ annihilation. The first three diagrams of Figure \ref{topprod} are present in the SM as well, the fourth diagram which is absent in the SM represents the effective $t\bar{t}gg$ vertex and is the expandable SM. It is also worthwhile to mention that as the semileptonic decay of the top (anti-top) takes place due to weak-interactions, the branching ratio of the top-quark will remain intact as of the SM.

At first, we start our calculation with the T-odd correlations induced by anomalous top-quark couplings defined in the following equations:
%%%%%%%%%%%%%%%%%%%%%%%%%%%%%%%%%%%%%%%%%%%%%%%%%%%%%%%%%%%%%%%%%%%%%
\begin{eqnarray}
\label{oldobs}
\nonumber
\mathcal C_1 &=&\epsilon(p_b,p_{\bar{b}},p_{l^+},p_{l^-})\\
\nonumber
\mathcal C_2 &=& \tilde 
q \cdot (p_{l^+}-p_{l^-})~\epsilon(p_{l^+},p_{l^-},p_{b}+p_{\bar{b}},\tilde q)\\
\nonumber
\mathcal C_3 &=& \tilde q \cdot (p_{l^+}-p_{l^-})~ 
\epsilon(p_b,p_{\bar{b}},p_{l^+}+p_{l^-},\tilde q)\\ 
\nonumber
\mathcal C_4 &=& \epsilon(P,p_b-p_{\bar{b}},p_{l^+},p_{l^-})\\ 
\mathcal C_5 &=& \epsilon(p_b + p_{l^+},p_{\bar{b}} + p_{l^{-}},p_b+p_{\bar{b}},p_{l^+}-p_{l^-}), 
\end{eqnarray}
%%%%%%%%%%%%%%%%%%%%%%%%%%%%%%%%%%%%%%%%%%%%%%%%%%%%%%%%%%%%%%%%%%%%%%  

where in the above expressions $\epsilon(a,b,c,d)=\epsilon_{\mu\nu\alpha\beta}a^{\mu}b^{\nu}c^{\alpha}d^{\beta}$ with $\epsilon_{\mu\nu\alpha\beta}$ being the Levi-Civita symbol of rank 4 which is completely anti-symmetric with $\epsilon_{0123}$~=~1 and $p_{b}~(p_{\bar{b}})$, $p_{l^+}~(p_{l^-})$ represent the four-momenta of $b~(\bar{b})$-quark, lepton (anti-lepton) respectively. P is the sum of four-momenta of b-quark, lepton, anti-b-quark and anti-lepton and $\tilde q$ is the difference of two-beam four momenta, defined as

%%%%%%%%%%%%%%%%%%%%%%%%%%%%%%%%%%%%%%%%%%%%%%%%%%%%%%%%%%%%%%%%%%%%
\begin{eqnarray}
\label{PQ eqn}
\nonumber
P &=& p_b + p_{l^{+}} + p_{\bar{b}} + p_{l^{-}}\\
\tilde q &=& P_1 - P_2.
\end{eqnarray}
%%%%%%%%%%%%%%%%%%%%%%%%%%%%%%%%%%%%%%%%%%%%%%%%%%%%%%%%%%%%%%%%%%%%%

It is interesting to note that the aforementioned observables neither require reconstruction of the produced tops nor any information about the spin of the produced particles. Also, a b-jet is distinguished with a $\bar{b}$-jet by measuring the direction of leptons i.e. the b-jet closer to $l^+$ is identified as the one arising due to a b-quark whereas the other b-jet closer to $l^{-}$ is identified as the one arising due to $\bar{b}$-quark.

%%%%%%%%%%%%%%%%%%%%%%%%%%%%%%%%%%%%%%%%%%%%%%%%%%%%%%%%%%%%%%%%%%%%%
%%%%%%%%%%%%%%%%%%%%%%%%%%%%%%%%%%%%%%%%%%%%%%%%%%%%%%%%%%%%%%%%%%%%%

Let us now consider observable $\mathcal C_1$ to check its CP properties \cite{Gupta:2009eq,Valencia:2013yr}

%%%%%%%%%%%%%%%%%%%%%%%%%%%%%%%%5%%%%%%%%%%%%%%%%%%%%%%%%%%%%%%%%%%%%%
\begin{eqnarray}
\label{bbframe}
\scalebox{0.80}{$
       \begin{aligned}
\nonumber
\mathcal C_1 &=& \epsilon(P_b,P_{\bar{b}},P_{l^+},P_{l^-}) \xrightarrow{\text{$(b\bar{b})_{CM}$}}~\propto\ \vec P_b \cdot (\vec P_{l^+}\times 
\vec P_{l^-})\\
\nonumber
 {\rm Now}~~~~\vec P_b \cdot (\vec P_{l^+}\times \vec P_{l^-})&\xrightarrow{\text C}& 
 \vec P_{\bar{b}} \cdot (\vec P_{l^-} \times \vec P_{l^+}) = -\vec P_{\bar{b}} \cdot (\vec 
P_{l^+} \times \vec P_{l^-}) =  \vec P_b \cdot (\vec P_{l^+} \times \vec P_{l^-})\\
 \vec P_b \cdot (\vec P_{l^+}\times \vec P_{l^-})&\xrightarrow{\text P}& -\vec 
P_b \cdot (-\vec P_{l^+}\times -\vec P_{l^-}) = -\vec P_b \cdot (\vec P_{l^+}\times \vec 
P_{l^-}).
\end{aligned}$}\\
\end{eqnarray}
%%%%%%%%%%%%%%%%%%%%%%%%%%%%%%%%%%%%%%%%%%%%%%%%%%%%%%%%%%%%%%%%%%%%%

In the above equation the left-hand side of the arrow describes the frame independent correlation and the right-hand side represents the correlation in a particular C.M. reference frame. In the first line of Eq. \ref{bbframe}, we go through $b\bar{b}$ C.M. frame which results in the triple product form. The obtained triple-product undergoes the charge conjugation and parity operation in the second and third lines, respectively, to ensure that it is $CP$-odd. Similarly, if we consider ($l^+l^-$) C.M. frame, the above observable takes the following form \cite{Gupta:2009eq,Valencia:2013yr}: 

%%%%%%%%%%%%%%%%%%%%%%%%%%%%%%%%%%%%%%%%%%%%%%%%%%%%%%%%%%%%%%%%%%%%%
\begin{eqnarray}
\label{llframe}
\scalebox{0.80}{$
   \begin{aligned}
\nonumber
\mathcal C_1 &=& \epsilon(P_b,P_{\bar{b}},P_{l^+},P_{l^-}) 
\xrightarrow{\text{$(l^+l^-)_{CM}$}}~\propto\ \vec P_{l^+} \cdot (\vec P_b\times \vec 
P_{\bar{b}})\\
\nonumber
{\rm Now}~~~~ \vec P_{l^+} \cdot (\vec P_b\times \vec 
P_{\bar{b}})&\xrightarrow{\text C}&\vec P_{l^-} \cdot (\vec P_{\bar{b}}\times \vec 
P_b) = -\vec P_{l^-} \cdot (\vec P_b\times \vec P_{\bar{b}}) = \vec P_{l^+} \cdot (\vec 
P_b\times \vec P_{\bar{b}})\\
\vec P_{l^+} \cdot (\vec P_b\times \vec P_{\bar{b}})&\xrightarrow{\text P}& -\vec 
P_{l^+} \cdot (-\vec P_b\times -\vec P_{\bar{b}}) = -\vec P_{l^+} \cdot (\vec P_b\times 
\vec P_{\bar{b}}).
\end{aligned}$}\\
\end{eqnarray}
%%%%%%%%%%%%%%%%%%%%%%%%%%%%%%%%%%%%%%%%%%%%%%%%%%%%%%%%%%%%

This further suggests that $\mathcal C_1$ is indeed CP-odd. In addition to the observables discussed in Eqs. \ref{oldobs}, we also construct the following new observables:

 %%%%%%%%%%%%%%%%%%%%%%%%%%%%%%%%%%%%%%%%%%%%%%%%%%%%%%%%%%%
\begin{eqnarray}
\label{newobs}
\nonumber
\mathcal C_6 &=& \epsilon(P,\tilde q,p_{l^+}+p_b,p_{l^-}+p_{\bar{b}})\\
\nonumber
\mathcal C_7 &=& \epsilon(P,\tilde q,p_{l^+},p_{l^-})\\
\mathcal C_8 &=& \epsilon(P,\tilde q,p_b,p_{\bar{b}}).
\end{eqnarray}    
%%%%%%%%%%%%%%%%%%%%%%%%%%%%%%%%%%%%%%%%%%%%%%%%%%%%%%%%%%%
\\
The advantage of considering these additional observables lies in the fact that these require lesser information than the observables defined in Eqs. \ref{oldobs}. For example, observable $\mathcal C_6$ requires information regarding the beam direction, a lepton having a positive charge and the associated b-quark and identifying a lepton having a negative charge and the associated anti-b-quark. Observable $\mathcal C_7$ requires information of the beam direction and leptons having a positive and negative charge. Similarly observable 
$\mathcal C_8$ requires information of the beam direction, b-quark and anti-b quark. In the next section, we will discuss the numerical simulation in detail.   
%%%%%%%%%%%%%%%%%%%%%%%%%%%%%%%%%%%%%%%%%%%%%%%%%%%%%%%%%%%
%%%%%%%%%%%%%%%%%%%%%%%%%%%%%%%%%%%%%%%%%%%%%%%%%%%%%%%%%%%

\section{Numerical Analysis} \label{Numanl}

In order to perform our study, we first produced $t\bar{t}$ pairs through the process $pp \to t\bar{t}$ and allowed them to decay semileptonically into $(bl^+\nu_{l})(\bar{b}l^-\bar{\nu_{l}})$ subsequently with the aid of $\tt MadGraph5$ \cite{Stelzer:1994ta, Alwall:2007st, Alwall:2008pm} at Leading order (LO) using the decay chain feature described in Ref. \cite{Alwall:2008pm}. Later these events are interfaced to $\tt Pythia8$ \cite{Sjostrand:2014zea} for Showering Hadronisation. The $CP$-violating interactions discussed in Eqs. \ref{oldobs} and \ref{newobs} have been incorporated in the $\tt MadGraph5$ via incorporating the Lagrangian given in Eq. \ref{intlag} in $\tt FeynRules$ \cite{Ask:2012sm}. The events are generated with the following selection criteria:
\\
%%%%%%%%%%%%%%%%%%%%%%%%%%%%%%%%%%%%%%%%%%%%%%%%%%%%%%%%%%%%
\begin{eqnarray}
\label{cuts}
P_T(l^\pm)& > & 20~{\rm GeV},
\nonumber\\
P_T(b,\bar{b})& > &25~{\rm GeV},
\nonumber\\
\eta(b,\bar{b},l^\pm)&<& 2.5, 
\nonumber\\
\Delta R (b\bar{b})& > & 0.4,
\nonumber\\
\Delta R (l^{+}l^{-})& > & 0.2,
\nonumber\\
\Delta R (bl)& > & 0.4,
\nonumber\\
{\not}E_{T} & > & 30~{\rm GeV}.
\end{eqnarray}
%%%%%%%%%%%%%%%%%%%%%%%%%%%%%%%%%%%%%%%%%%%%%%%%%%%%%%%%%%%

The experimental values of the input parameters considered in our study are presented in Table \ref{SMinputs}, the renormalisation and factorisation scale has been set to 91.188 GeV and the parton distribution functions had been considered to be nn23lo1 \cite{Ball:2013hta, Ball:2014uwa}.

%%%%%%%%%%%%%%%%%%%%%%%%%%%%%%%%%%%%%%%%%%%%%%%%%%%%%%%%%%%%
\begin{table}[!ht]
\centering
\scalebox{1.2}{
%\begin{adjustbox}{width=1\textwidth}
%\small
\renewcommand{\arraystretch}{1.4}
\begin{tabular} {|c|c|}
\hline
SM parameter & Experimental value\\
\hline
$m_b (m_b)$ & 4.7 $\pm$ 0.06 GeV\\
$m_t(m_t)$ & 173.0 $\pm$ 0.4 GeV\\
$M_W$  & 80.387 $\pm$ 0.02 GeV \\
$\alpha_s^{\overline {MS}}(M_z)$ & 0.118 $\pm$ 0.001\\ 
 \hline
\end{tabular}}
\caption{Experimental values of Standard Model input parameters \cite{Tanabashi:2018oca}.}
\label{SMinputs}
\end{table}
%%%%%%%%%%%%%%%%%%%%%%%%%%%%%%%%%%%%%%%%%%%%%%%%%%%%%%%%%%%%
\FloatBarrier

In order to estimate the asymmetries at the LHC, we generate $pp \to t\bar{t} \to (bl^+\nu_{l})(\bar{b}l^-\bar{\nu_{l}})$ events with the aid of $\tt MadGraph5$ at 13 TeV and 14 TeV LHC energies with distinctive values of coupling constant ($d_g$) and scale parameter ($\Lambda$) for the observables given in Eqs. \ref{oldobs} and \ref{newobs}. The values of coupling constant $d_g$ and scale parameter $\Lambda$ have been considered from 0 to 5$\times10^{-2}$ and $M_W$ to 2 TeV respectively where $d_g$ = 0 is actually SM. The associated CP-violating asymmetry for the observables listed in Eqs. \ref{oldobs} and \ref{newobs} is constructed using the formula:
%%%%%%%%%%%%%%%%%%%%%%%%%%%%%%%%%%%%%%%%%%%%%%%%%%%%%%%%%%%%
\begin{eqnarray}
\label{Asymmetry formula}
\mathcal A_{CP} = {\frac{N(\mathcal C_i>0) - N(\mathcal C_i<0)}{N(\mathcal 
C_i>0) + N(\mathcal C_i<0)}},
\end{eqnarray}
%%%%%%%%%%%%%%%%%%%%%%%%%%%%%%%%%%%%%%%%%%%%%%%%%%%%%%%%%%%%

where the numerator represents the difference between the number of events having positive and negative values of the observable whereas the denominator represents the total number of events. Clearly, for a CP-symmetric observable, $\mathcal A_{CP}$ would be zero because the number of events with a positive value of observable will be equal to the number of events with a negative value of observable and non-zero otherwise. The number of experimentally measured $pp \to t\bar{t} \to (bl^+\nu_{l})(\bar{b}l^-\bar{\nu_{l}})$ events at the LHC are given by 

%%%%%%%%%%%%%%%%%%%%%%%%%%%%%%%%%%%%%%%%%%%%%%%%%%%%%%%%%%
\begin{eqnarray}
\label{sensitivity}
N^{exp} &=& \sigma^{exp} \times {\rm Br}(t \to bl\nu)^2\times (b_{\rm{tag}})^2\times 
\epsilon_{\rm{eff}}\times \int \mathcal Ldt,
\end{eqnarray}
%%%%%%%%%%%%%%%%%%%%%%%%%%%%%%%%%%%%%%%%%%%%%%%%%%%%%%%%%%

where $\sigma^{exp}$ represents the experimentally measured value of the $t\bar{t}$ cross-section at a given C.M. energy at the LHC, $b_{\rm{tag}}$ is the b-tagging efficiency, $\epsilon_{\rm{eff}}$ is the efficiency of cuts and $\int \mathcal L dt$ represents the integrated luminosity at the LHC. The sensitivity for a given observable could be estimated by comparing the $\mathcal A_{CP}$ corresponding to the underlying observable with the following experimental sensitivity at a given confidence level (C.L.) $n_{cl}$:
\\
%%%%%%%%%%%%%%%%%%%%%%%%%%%%%%%%%%%%%%
\begin{eqnarray}
\centering
\label{expectation}
\mathcal A^{exp} &=& {\frac{n_{cl}}{\sqrt{N^{exp}}}},    
\end{eqnarray} 
%%%%%%%%%%%%%%%%%%%%%%%%%%%%%%%%%%%%%%

\begin{figure*}[h!]
\begin{tabular}{c c}
\hspace{-0.3cm}
\includegraphics[width=0.49\textwidth,height=4.9cm]{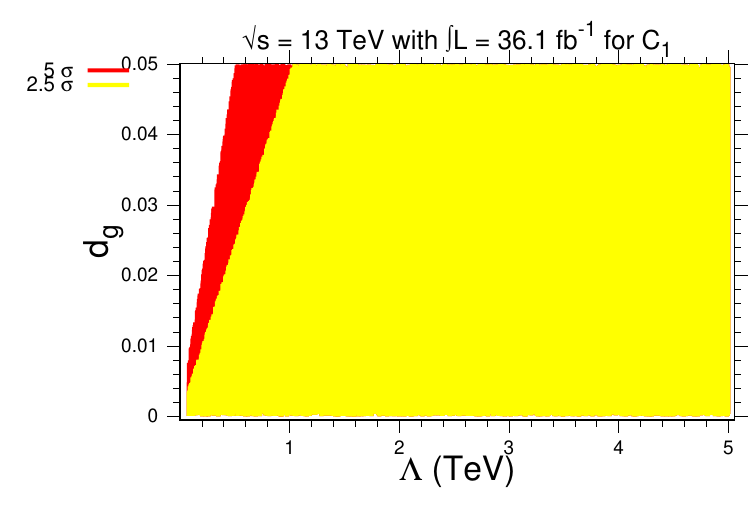}\hspace{-0.3cm}
&\includegraphics[width=0.49\textwidth,height=4.9cm]{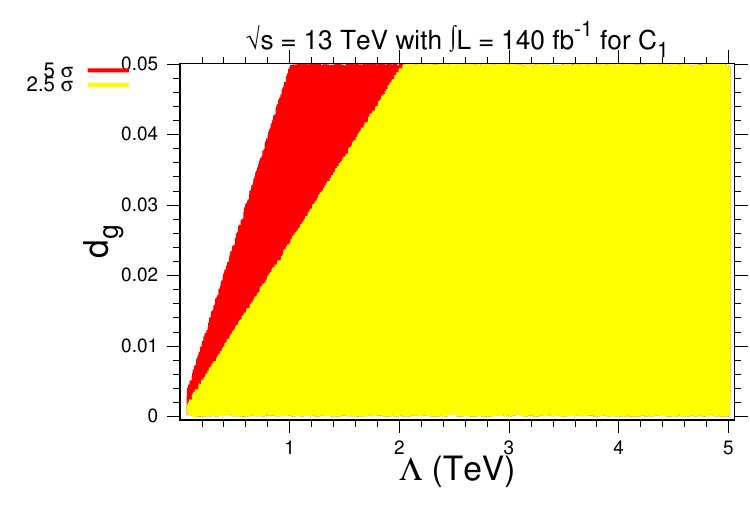}\hspace{-0.3cm}
\end{tabular}
 \caption{$d_g$ vs. $\Lambda$ for observable $\mathcal C_1$ at $\sqrt {S}$ = 13 TeV energy at LHC for an integrated luminosity of (a) 36.1 fb$^{-1}$ and (b) 140 fb$^{-1}$ respectively.}
 \label{Plot13TeVC1}
\end{figure*}

\begin{figure*}[h!]
\begin{tabular}{c c}
\hspace{-0.3cm}
\includegraphics[width=0.49\textwidth,height=4.9cm]{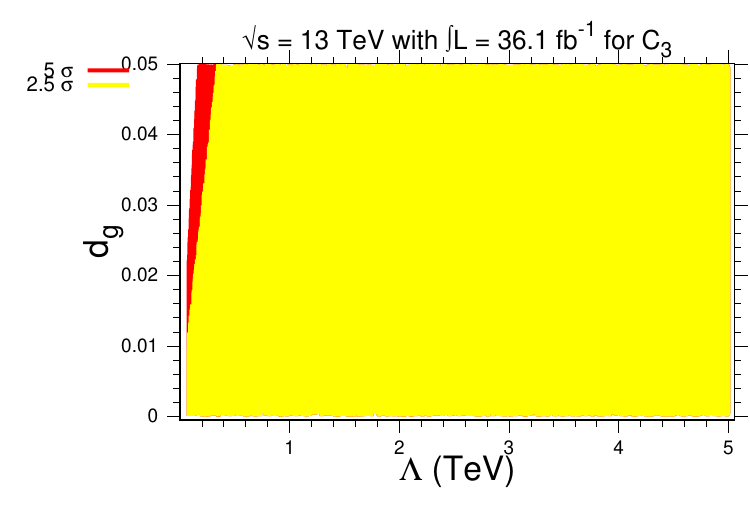}\hspace{-0.3cm}
&\includegraphics[width=0.49\textwidth,height=4.9cm]{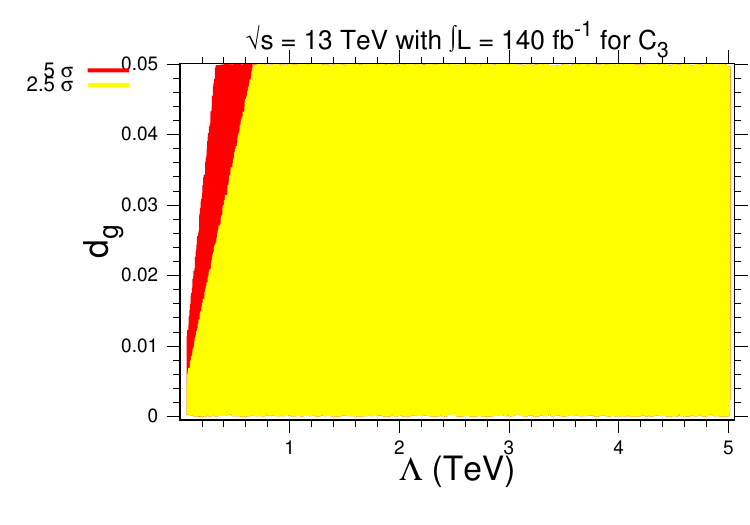}\hspace{-0.3cm}
\end{tabular}
 \caption{$d_g$ vs. $\Lambda$ for observable $\mathcal C_3$ at $\sqrt {S}$ = 13 TeV energy at LHC for an integrated luminosity of (a) 36.1 fb$^{-1}$ and (b) 140 fb$^{-1}$ respectively.}
 \label{Plot13TeVC3}
\end{figure*}

\begin{figure*}[h!]
\begin{tabular}{c c}
\hspace{-0.3cm}
\includegraphics[width=0.49\textwidth,height=4.9cm]{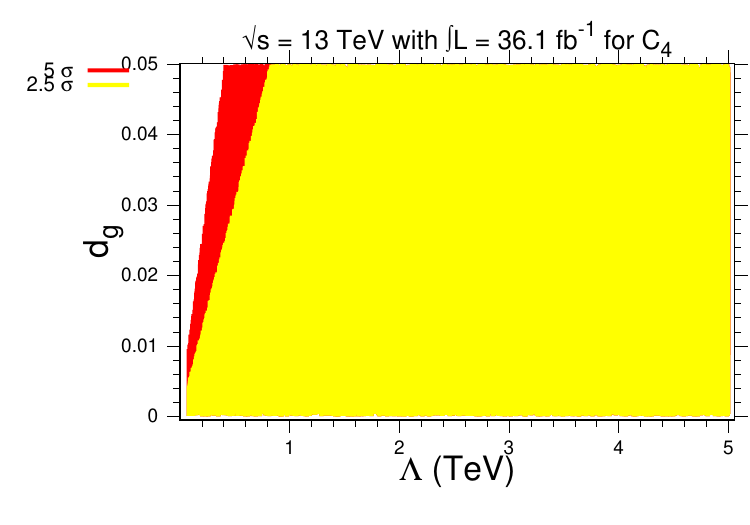}\hspace{-0.3cm}
&\includegraphics[width=0.49\textwidth,height=4.9cm]{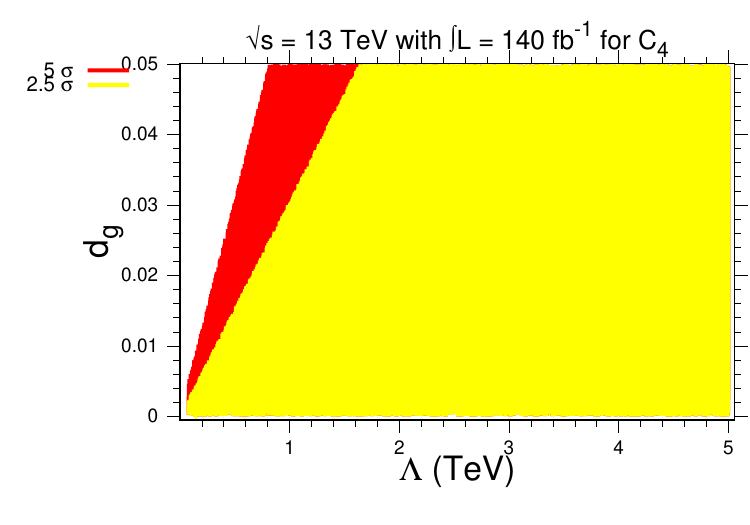}\hspace{-0.3cm}
\end{tabular}
 \caption{$d_g$ vs. $\Lambda$ for observable $\mathcal C_4$ at $\sqrt {S}$ = 13 TeV energy at LHC for an integrated luminosity of (a) 36.1 fb$^{-1}$ and (b) 140 fb$^{-1}$ respectively.}
 \label{Plot13TeVC4}
\end{figure*}

\begin{figure*}[h!]
\begin{tabular}{c c}
\hspace{-0.3cm}
\includegraphics[width=0.49\textwidth,height=4.9cm]{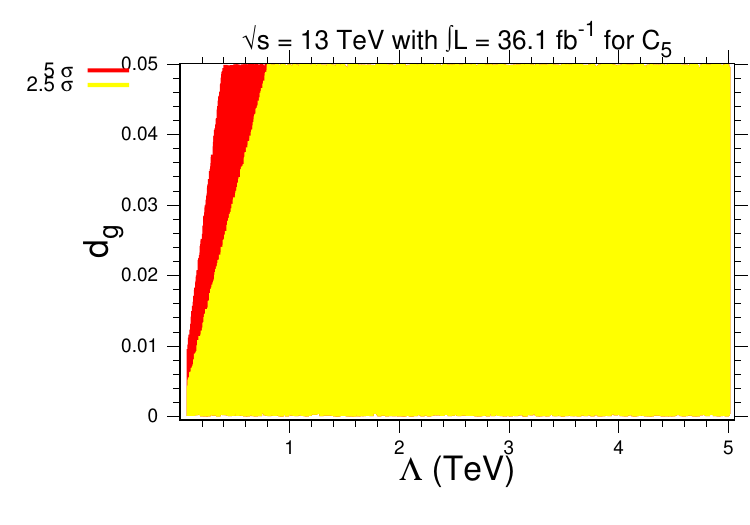}\hspace{-0.3cm}
&\includegraphics[width=0.49\textwidth,height=4.9cm]{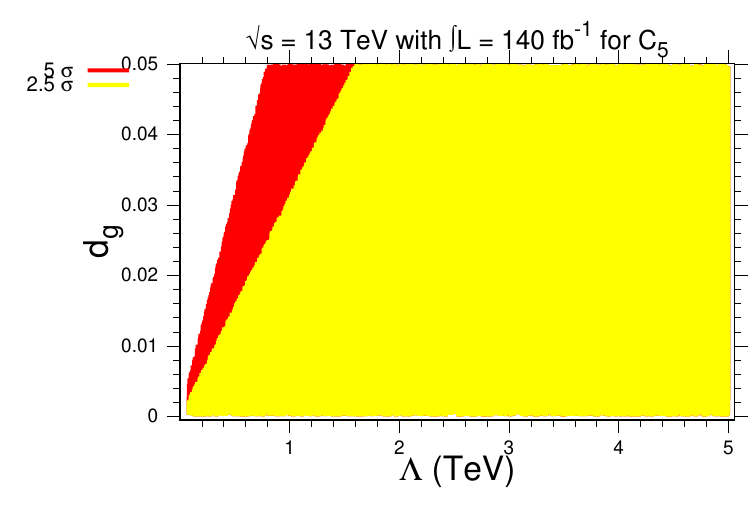}\hspace{-0.3cm}
\end{tabular}
 \caption{$d_g$ vs. $\Lambda$ for observable $\mathcal C_5$ at $\sqrt {S}$ = 13 TeV energy at LHC for an integrated luminosity of (a) 36.1 fb$^{-1}$ and (b) 140 fb$^{-1}$ respectively.}
 \label{Plot13TeVC5}
\end{figure*}

\begin{figure*}[h!]
\begin{tabular}{c c}
\hspace{-0.3cm}
\includegraphics[width=0.49\textwidth,height=5.8cm]{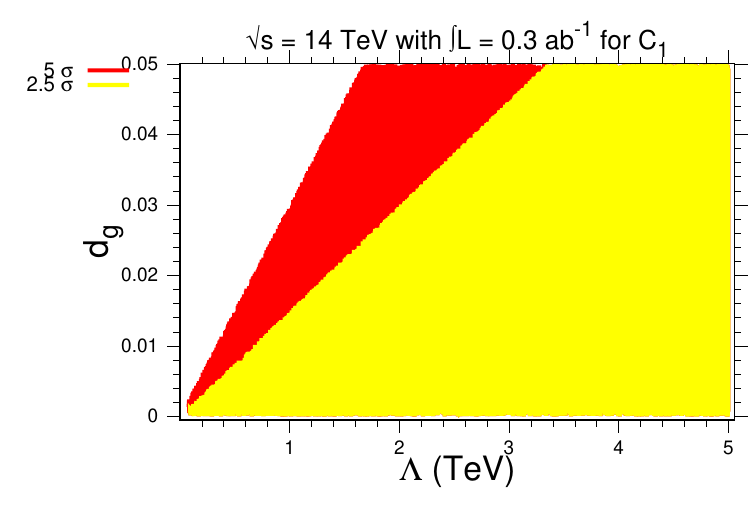}\hspace{-0.3cm}
&\includegraphics[width=0.49\textwidth,height=5.8cm]{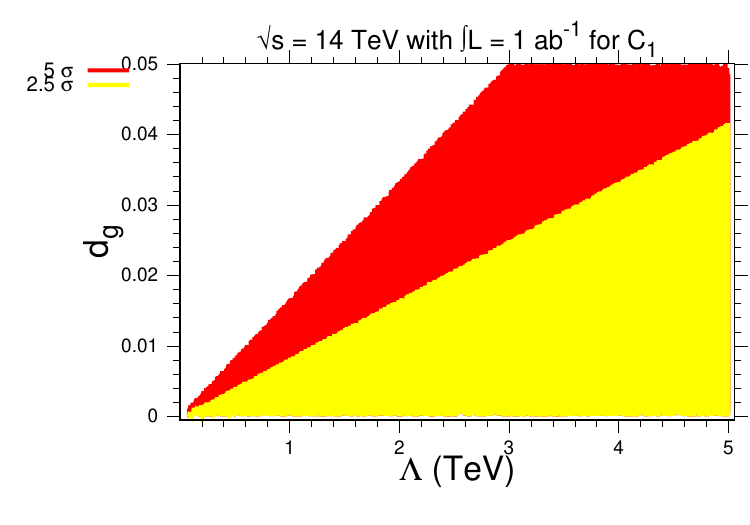}\hspace{-0.3cm}
\\
\hspace{-0.3cm}
\includegraphics[width=0.49\textwidth,height=5.8cm]{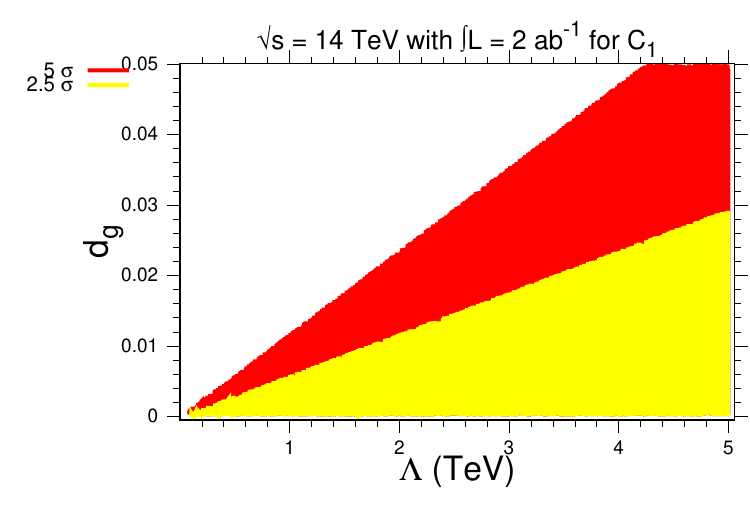}\hspace{-0.3cm}
&\includegraphics[width=0.49\textwidth,height=5.8cm]{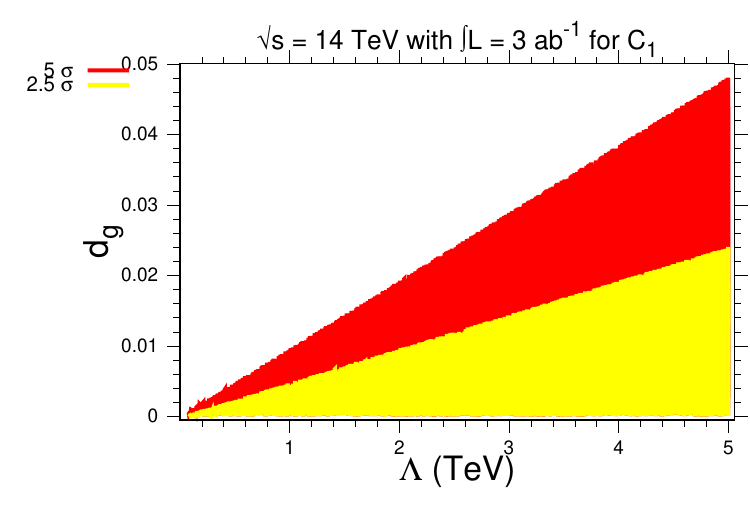}\hspace{-0.3cm}
\end{tabular}
 \caption{$d_g$ vs. $\Lambda$ for observable $\mathcal C_1$ at $\sqrt {S}$ = 14 TeV 
energy at LHC for an integrated luminosity of (a) 0.3 ab$^{-1}$, (b) 1 ab$^{-1}$, (c) 2 ab$^{-1}$ and (d) 3 ab$^{-1}$ respectively.}
 \label{Plot14TeVC1}
\end{figure*}

\begin{figure*}[h!]
\begin{tabular}{c c}
\hspace{-0.3cm}
\includegraphics[width=0.49\textwidth,height=5.8cm]{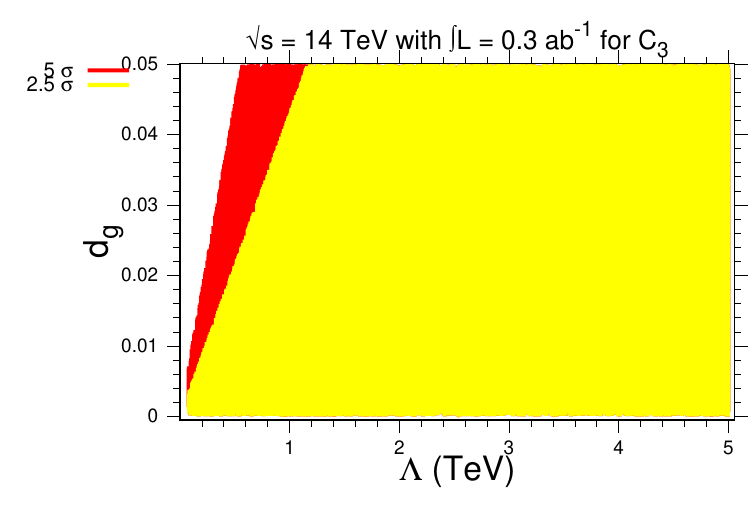}\hspace{-0.3cm}
&\includegraphics[width=0.49\textwidth,height=5.8cm]{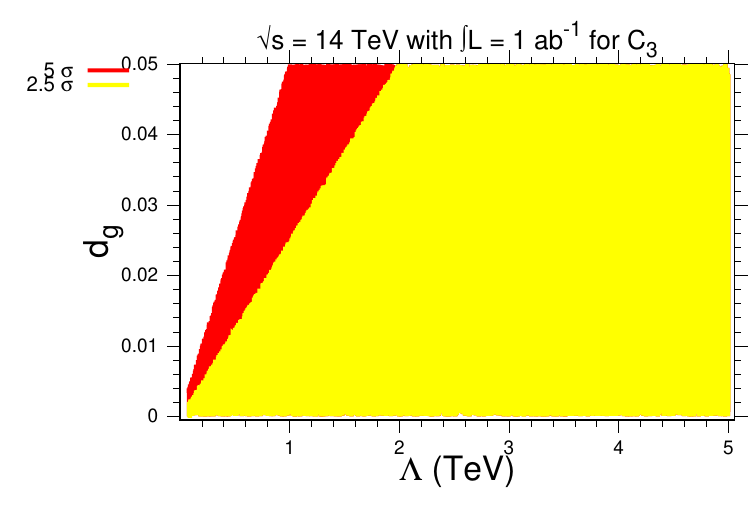}\hspace{-0.3cm}
\\
\hspace{-0.3cm}
\includegraphics[width=0.49\textwidth,height=5.8cm]{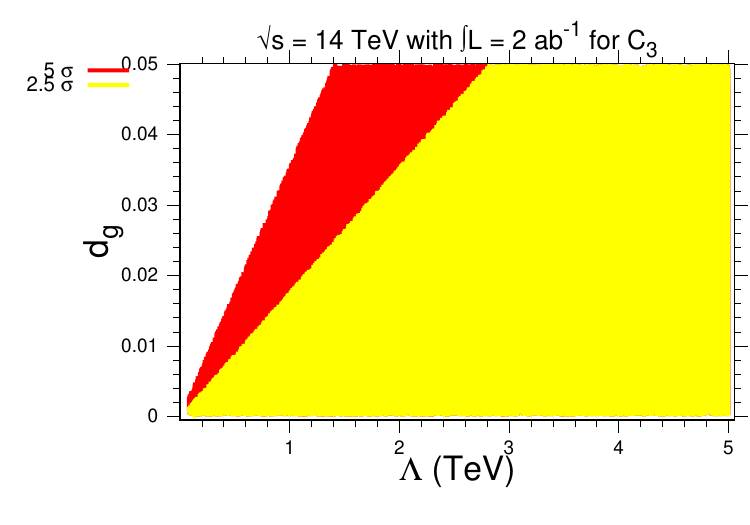}\hspace{-0.3cm}
&\includegraphics[width=0.49\textwidth,height=5.8cm]{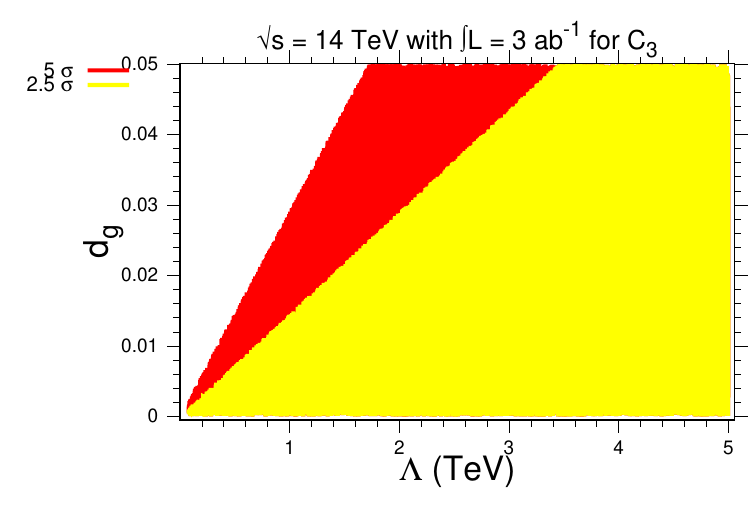}\hspace{-0.3cm}
\end{tabular}
 \caption{$d_g$ vs. $\Lambda$ for observable $\mathcal C_3$ at $\sqrt {S}$ = 14 TeV 
energy at LHC for an integrated luminosity of (a) 0.3 ab$^{-1}$, (b) 1 ab$^{-1}$, (c) 2 ab$^{-1}$ and (d) 3 ab$^{-1}$ respectively.}
 \label{Plot14TeVC3}
\end{figure*}

\begin{figure*}[h!]
\begin{tabular}{c c}
\hspace{-0.3cm}
\includegraphics[width=0.49\textwidth,height=5.8cm]{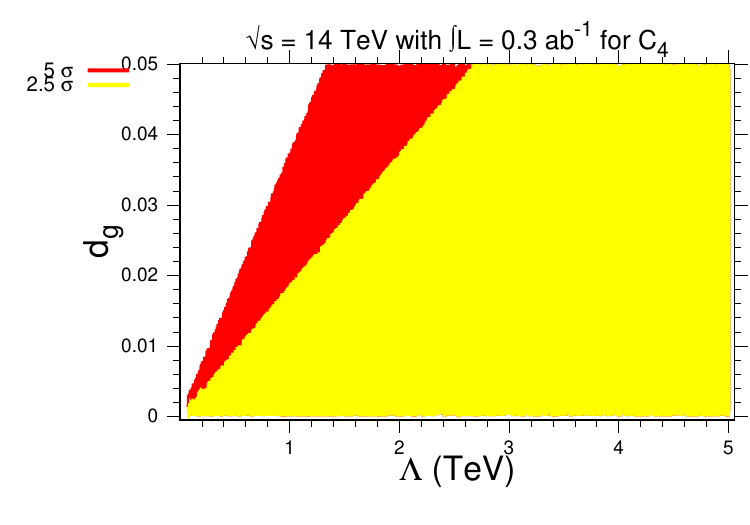}\hspace{-0.3cm}
&\includegraphics[width=0.49\textwidth,height=5.8cm]{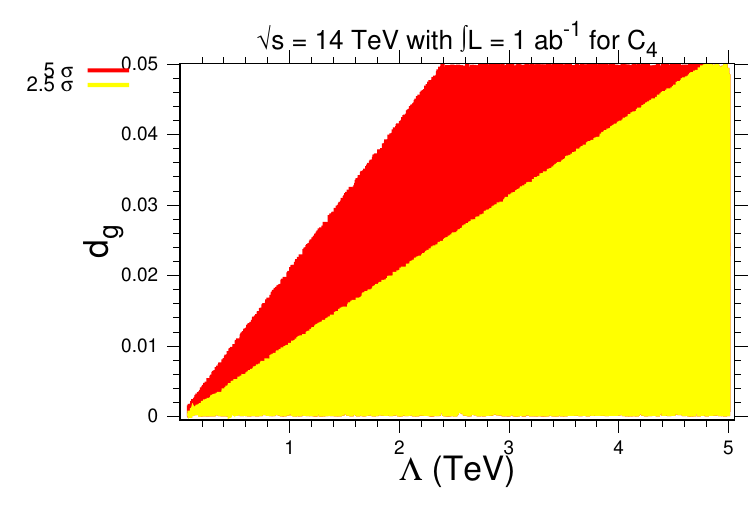}\hspace{-0.3cm}
\\
\hspace{-0.3cm}
\includegraphics[width=0.49\textwidth,height=5.8cm]{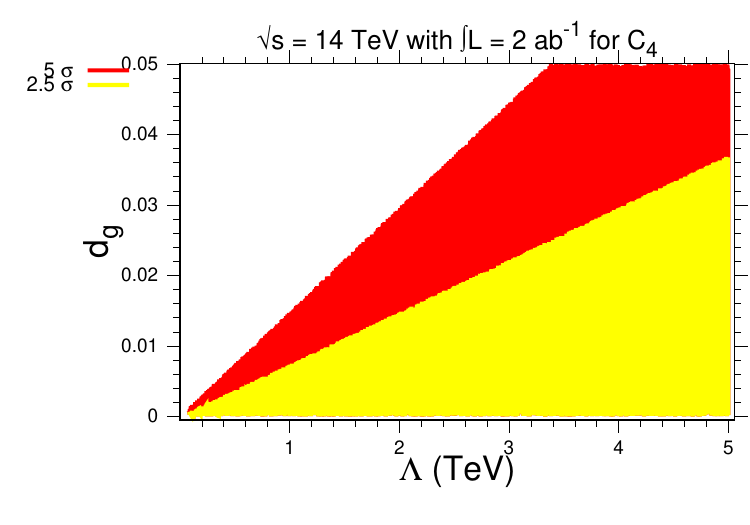}\hspace{-0.3cm}
&\includegraphics[width=0.49\textwidth,height=5.8cm]{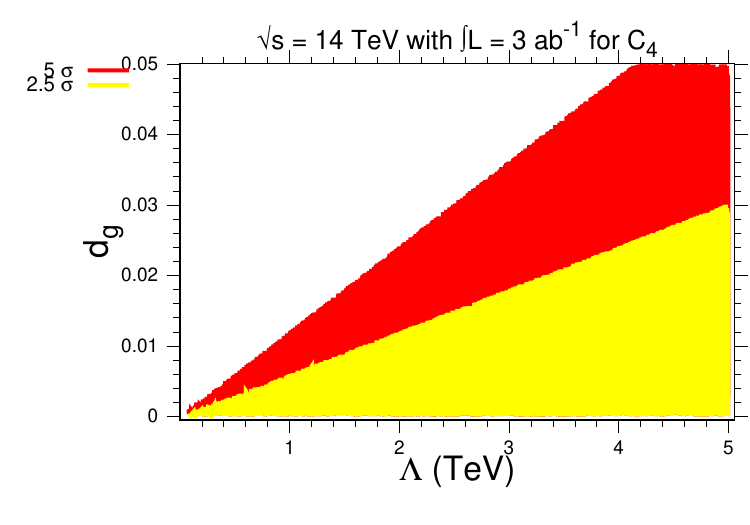}\hspace{-0.3cm}
\end{tabular}
 \caption{$d_g$ vs. $\Lambda$ for observable $\mathcal C_4$ at $\sqrt {S}$ = 14 TeV 
energy at LHC for an integrated luminosity of (a) 0.3 ab$^{-1}$, (b) 1 ab$^{-1}$, (c) 2 ab$^{-1}$ and (d) 3 ab$^{-1}$ respectively.} 
\label{Plot14TeVC4}
\end{figure*}

\begin{figure*}[h!]
\begin{tabular}{c c}
\hspace{-0.3cm}
\includegraphics[width=0.49\textwidth,height=5.8cm]{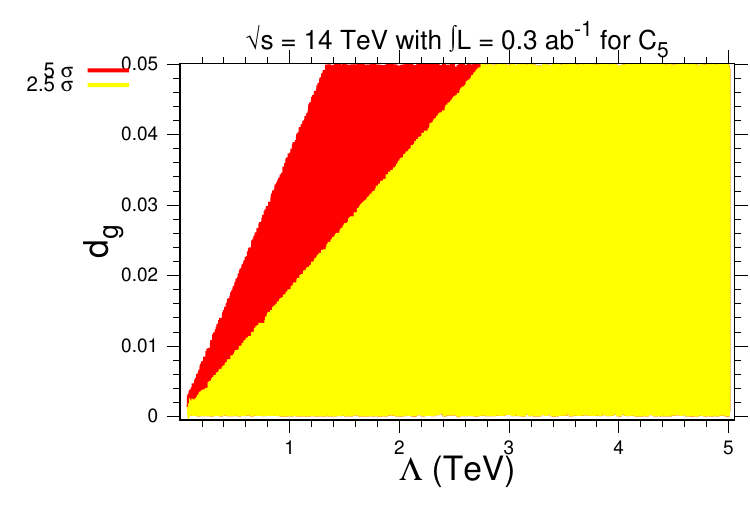}\hspace{-0.3cm}
&\includegraphics[width=0.49\textwidth,height=5.8cm]{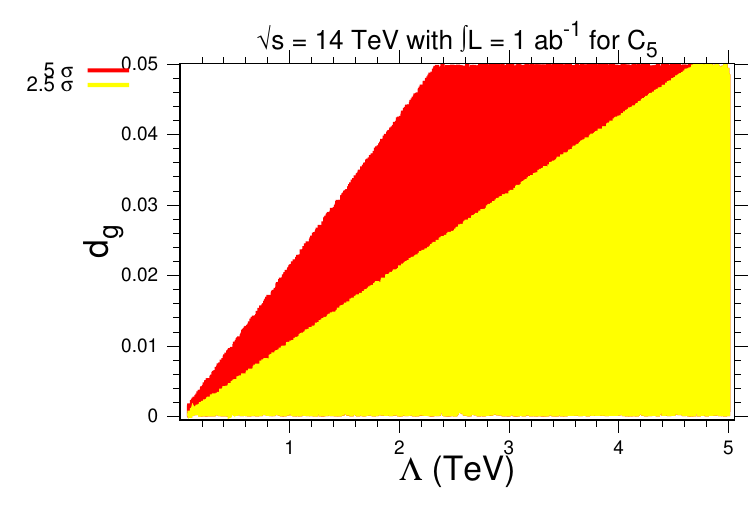}\hspace{-0.3cm}
\\
\hspace{-0.3cm}
\includegraphics[width=0.49\textwidth,height=5.8cm]{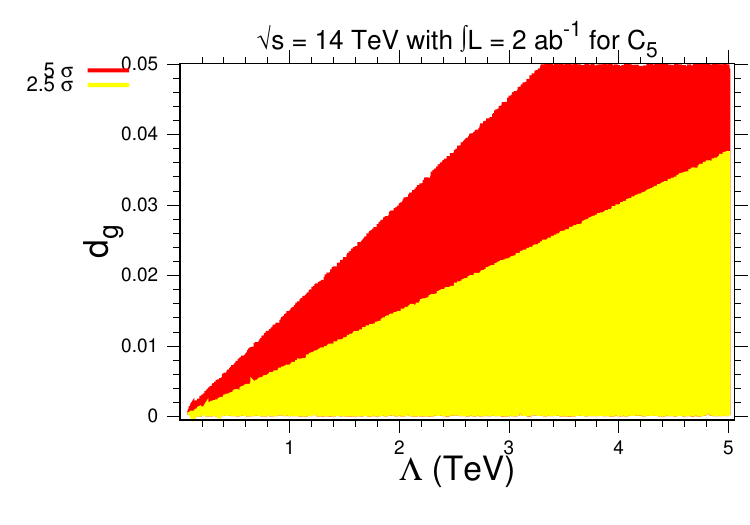}\hspace{-0.5cm}
&\includegraphics[width=0.49\textwidth,height=5.8cm]{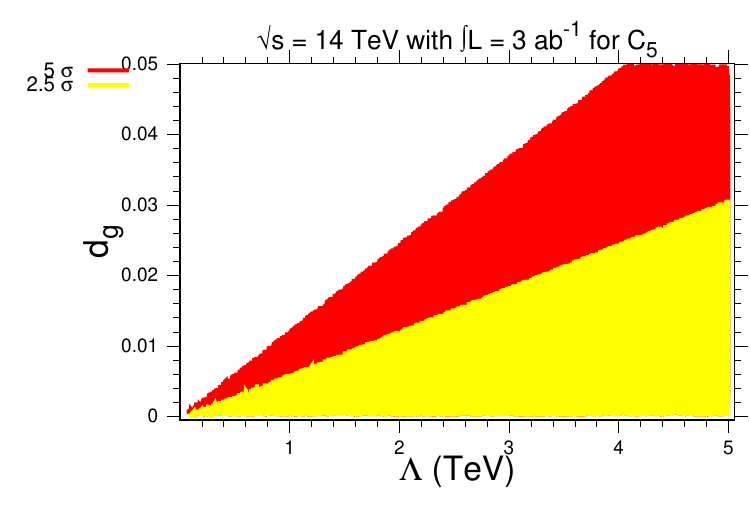}\hspace{-0.5cm}
\end{tabular}
 \caption{$d_g$ vs. $\Lambda$ for observable $\mathcal C_5$ at $\sqrt {S}$ = 14 TeV 
energy at LHC for an integrated luminosity of (a) 0.3 ab$^{-1}$, (b) 1 ab$^{-1}$, (c) 2 ab$^{-1}$ and (d) 3 ab$^{-1}$ respectively.}
\label{Plot14TeVC5} 
\end{figure*}

These are discussed in Figures \ref{Plot13TeVC1} -- \ref{Plot14TeVC5} for $\sqrt{S}$ = 13 TeV and 14 TeV at the LHC. The values of asymmetries corresponding to various CP-violating observables discussed in Eqs. \ref{oldobs} and \ref{newobs} are also presented for various values of $\Lambda$ and $d_g$. We estimate asymmetries for $d_g$ from 0 to 0.05 and $\Lambda$ between $M_W$ to 2 TeV for $\sqrt{S}$ = 13 TeV and 14 TeV at the LHC. In Tables \ref{results13TeV} and \ref{results14TeV}, we present asymmetries corresponding to various observables at $\sqrt{S}$ = 13 TeV and 14 TeV LHC energies. From these tables, it is clear that the asymmetries corresponding to observables $C_2,~C_6,~C_7$ and $C_8$ are within the limits of statistical uncertainties and therefore would not be useful to calculate CP-violation sensitivity as these are consistent with SM. However, asymmetries related to observables $\mathcal C_1,~ \mathcal C_3,~\mathcal C_4$ and $\mathcal C_5$ are found to be non-zero at 3$\sigma$ C.L.. It is, therefore, informative to discuss the asymmetries obtained for observables $\mathcal C_1,~\mathcal C_3,~\mathcal C_4$ and $\mathcal C_5$ in detail as these are expected to be more sensitive.    

%%%%%%%%%%%%%%%%%%%%%%%%%%%%%%%%%%%%%%%%%%%%%%%%%%%%%%%%%%%%
\begin{table}[!ht]
\centering
\scalebox{1.2}{
%\begin{adjustbox}{width=1\textwidth}
%\small
\renewcommand{\arraystretch}{1.4}
\begin{tabular} {|c|c|c|c|c|c|c|c|c|c|}
\hline
$\Lambda$ & ${d_g}$ & $\mathcal{A}_1$ & $\mathcal{A}_2$ & $\mathcal{A}_3$ & $\mathcal{A}_4$ & $\mathcal{A}_5$ & $\mathcal{A}_6$ & $\mathcal{A}_7$ & $\mathcal{A}_8$\\
\hline
& SM & 0.05 & 0.01 & 0.02 & -0.05 & 0.03 & 0.05 & -0.05 & 0.04\\
\hline
 & 5$\times 10^{-3}$ & 1.17 & 0 & 0.37 & -0.91 & 0.89 & 0.02 & -0.01 & -0.01\\
M$_W$ & 1$\times 10^{-2}$ & 2.22 & 0 & 0.72 & -1.74 & 1.74 & 0.04 & -0.02 & 0.04\\
  & 5$\times 10^{-2}$ & 6.29 & 0.02 & 2.12 & -5.11 & 4.98 & 0.01 & -0.01 & 0.03\\
\hline
  & 5$\times 10^{-3}$ & 0.19 & 0.01 & 0.02 & -0.15 & 0.13 & -0.01 & 0.03 & -0.03\\
0.5 TeV & 1$\times 10^{-2}$ & 0.38 & 0.04 & 0.17 & -0.29 & 0.31 & -0.03 & 0.05 & -0.06\\
  & 5$\times 10^{-2}$ & 1.79 & -0.01 & 0.56 & -1.47 & 1.37 & -0.02 & 0.03 & -0.04\\
\hline
 & 5$\times 10^{-3}$ & 0.06 & 0.02 & 0.07 & -0.11 & 0.04 & -0.03 & 0.03 & -0.09\\
1 TeV & 1$\times 10^{-2}$ & 0.14 & -0.03 & 0.08 & -0.12 & 0.13 & -0.04 & 0.02 & 0\\
  & 5$\times 10^{-2}$ & 0.92 & 0 & 0.28 & -0.72 & 0.74 & -0.03 & 0 & 0.03\\
 \hline
 & 5$\times 10^{-3}$ & 0.03 & -0.01 & -0.01 & -0.01 & 0.04 & 0.01 & 0 & 0\\
2 TeV & 1$\times 10^{-2}$ & 0.12 & 0.02 & 0.03 & -0.10 & 0.10 & 0 & 0.01 & -0.03\\
  & 5$\times 10^{-2}$ & 0.42 & 0.02 & 0.13 & -0.33 & 0.32 & 0.02 & -0.03 & -0.04\\
 \hline
\end{tabular}}
\caption{Integrated asymmetries (in $\%$) at LO for $\sqrt S$ = 13 TeV at LHC for the 
process $pp \to t\bar{t} \to (bl^+\nu_l)(\bar{b}l^-\bar{\nu_l})$ corresponding to various observables for distinct choices of ${d_g}$ and $\Lambda$. The statistical uncertainty at the 1$\sigma$ confidence level in all the results is estimated to be about 3$\times 10^{-4}$.}
\label{results13TeV}
\end{table}
%%%%%%%%%%%%%%%%%%%%%%%%%%%%%%%%%%%%%%%%%%%%%%%%%%%%%%%%%%%%
\FloatBarrier

From Tables \ref{results13TeV} and \ref{results14TeV}, it is also clear that if we fix the CP-violating scale to a certain value the asymmetries increase linearly with $d_g$ which supports the results in Refs. \cite{Gupta:2009eq, Gupta:2009wu}. Conversely, limiting the coupling $d_g$ to a constant value and increasing the value of $\Lambda$ reduces the value of the resulting asymmetries. This suggests that large CP-violation sensitivity can be achieved in two ways, either increasing $d_g$ or decreasing $\Lambda$. Furthermore, the asymmetries obtained at the $\sqrt{S}$ = 14 TeV energy at LHC, presented in Table \ref{results14TeV}, show similar results as observed for the 13 TeV LHC energy. According to the above tables, we infer that the largest asymmetry corresponds to the observable $\mathcal C_1$. The results corresponding to non-zero asymmetry could also be summarised as

%%%%%%%%%%%%%%%%%%%%%%%%%%%%%%%%%%%%%%%%%%%%%%%%%%%%%%%%%%%%
\begin{eqnarray}
\label{funcform13TeV}
\mathcal A_1 &=& 0.0023 + 119.3~\frac{d_g}{\Lambda},
\nonumber\\
\mathcal A_3 &=& 0.0007 + 39.9~\frac{d_g}{\Lambda},
\nonumber\\
\mathcal A_4 &=& -0.0018 - 96.1~\frac{d_g}{\Lambda},
\nonumber\\
\mathcal A_5 &=& 0.0018 + 93.6~\frac{d_g}{\Lambda},
\end{eqnarray}
%%%%%%%%%%%%%%%%%%%%%%%%%%%%%%%%%%%%%%%%%%%%%%%%%%%%%%%%%%

respectively for observables $\mathcal C_1,~\mathcal C_3,~\mathcal C_4$ and $\mathcal C_5$. 

%%%%%%%%%%%%%%%%%%%%%%%%%%%%%%%%%%%%%%%%%%%%%%%%%%%%%%%%%%%%
\begin{table}[!ht]
\centering
\scalebox{1.2}{
%\begin{adjustbox}{width=1\textwidth}
%\small
\renewcommand{\arraystretch}{1.4}
\begin{tabular} {|c|c|c|c|c|c|c|c|c|c|}
\hline
$\Lambda$ & ${d_g}$ & $\mathcal{A}_1$ & $\mathcal{A}_2$ & $\mathcal{A}_3$ & $\mathcal{A}_4$ & $\mathcal{A}_5$ & $\mathcal{A}_6$ & $\mathcal{A}_7$ & $\mathcal{A}_8$\\
    \hline
 & SM & -0.03 & 0.02 & 0.01 & 0.07 & -0.04 & -0.03 & -0.03 & -0.03\\
\hline
 & 5$\times 10^{-3}$ & 1.12 & 0 & 0.36 & -0.95 & 0.88 & -0.03 & 0.01 & -0.01\\
M$_W$ & 1$\times 10^{-2}$ & 2.24 & 0.03 & 0.70 & -1.73 & 1.74 & -0.03 & -0.04 & -0.01\\
  & 5$\times 10^{-2}$ & 6.39 & -0.02 & 2.09 & -5.11 & 4.97 & 0.05 & -0.01 & -0.03\\
\hline
  & 5$\times 10^{-3}$ & 0.17 & -0.04 & 0.10 & -0.13 & 0.14 & 0.02 & 0.03 & -0.04\\
0.5 TeV & 1$\times 10^{-2}$ & 0.36 & 0.01 & 0.14 & -0.33 & 0.24 & 0.01 & 0.01 & -0.03\\
  & 5$\times 10^{-2}$ & 1.89 & -0.06 & 0.62 & -1.46 & 1.46 & 0.02 & 0.07 & -0.03\\
\hline
 & 5$\times 10^{-3}$ & 0.14 & 0.03 & 0.05 & -0.10 & 0.12 & 0.02 & -0.04 & -0.01\\
1 TeV & 1$\times 10^{-2}$ & 0.21 & 0.03 & 0.02 & -0.21 & 0.16 & -0.03 & 0.02 & 0\\
  & 5$\times 10^{-2}$ & 0.89 & 0.01 & 0.23 & -0.75 & 0.75 & 0 & 0.03 & -0.06\\
 \hline
 & 5$\times 10^{-3}$ & 0.07 & 0.01 & 0.03 & -0.03 & 0.09 & -0.03 & -0.05 & -0.01\\
2 TeV  & 1$\times 10^{-2}$ & 0.09 & -0.02 & 0.07 & -0.04 & 0.05 & -0.01 & 0 & 0\\
  & 5$\times 10^{-2}$ & 0.45 & -0.03 & 0.14 & -0.33 & 0.33 & 0 & -0.01 & 0.03\\
 \hline
\end{tabular}}
\caption{Integrated asymmetries (in $\%$) at LO for $\sqrt S$ = 14 TeV at LHC for the process $pp \to t\bar{t} \to (bl^+\nu_l)(\bar{b}l^-\bar{\nu_l})$ corresponding to various observables for distinct choices of ${d_g}$ and $\Lambda$. The statistical uncertainty at the 1$\sigma$ confidence level in all the results is estimated to be about 3$\times 10^{-4}$.}
\label{results14TeV}
\end{table}
%%%%%%%%%%%%%%%%%%%%%%%%%%%%%%%%%%%%%%%%%%%%%%%%%%%%%%%%%%%%
\FloatBarrier

It is to be noted that for estimating the experimental uncertainties in event rates we first combined the ATLAS \cite{TheATLAScollaboration:2016hcb} and CMS \cite{CMS:2016syx} experimental uncertainties observed with 2015 and 2016 data during LHC Run \RomanNumeralCaps{2} for the top pair at $\sqrt{S}$ = 13 TeV for 36.1 ${\rm fb^{-1}}$ presented in Ref. \cite{Mengarelli:2017rmu}. In order to calculate experimental sensitivity, we first combined the ATLAS and CMS cross-sections which are as follows:
%%%%%%%%%%%%%%%%%%%%%%%%%%%%%%%%%
\begin{eqnarray}
\label{LHC cross-section}
\sigma^{ATLAS} &=& {\rm 803 \pm 7(stat) \pm 27(Syst) \pm 45(lumi) \pm 
12(beam)~pb}, 
\nonumber\\
\sigma^{CMS} &=& {\rm 793 \pm 8(stat) \pm 38(Syst) \pm 21(lumi)~pb},
\nonumber\\
\sigma^{LHC} &=& {\rm 798 \pm 49.25~ pb}.
\end{eqnarray}
%%%%%%%%%%%%%%%%%%%%%%%%%%%%%%%%%
\\
Event rates were then estimated by combining the cross-section with the luminosity, branching ratios for the $t \to bl\nu_{_l}$ and the b-tagging efficiency which is assumed to be 56 $\%$. A similar calculation has been performed for $\sqrt{S}$ = 14 TeV with a theoretical cross-section $953.6_{-33.9-17.8}^{+22.7+16.2}$ pb at NNLO + NNLL level \cite{Czakon:2013goa} for the projected integrated luminosities of the LHC of $\int \mathcal L dt$ = 0.3 ab$^{-1}$, 1 ab$^{-1}$, 2 ab$^{-1}$ and 3 ab$^{-1}$. We show the results for 13 TeV and 14 TeV C.M. energies at LHC for ${d_g}$ vs. $\Lambda$ at various confidence levels in Figures \ref{Plot13TeVC1} -- \ref{Plot14TeVC5}. We present the results for $\Lambda$ between the range 0 to 5 TeV but we had actually performed the study between the range $M_W$ to 2 TeV.
\\
%%%%%%%%%%%%%%%%%%%%%%%%%%%%%%%%%%%%%%%%%%%%%%%%%%%%%%%%%%%%%%%%%%%
%%%%%%%%%%%%%%%%%%%%%%%%%%%%%%%%%%%%%%%%%%%%%%%%%%%%%%%%%%%%%%%%%%
%%%%%%%%%%%%%%%%%%%%%%%%%%%%%%%%%%%%%%%%%%%%%%%%%%%%%%%%%%%%%%%%%%%
%%%%%%%%%%%%%%%%%%%%%%%%%%%%%%%%%%%%%%%%%%%%%%%%%%%%%%%%%%%%%%%%%%
%%%%%%%%%%%%%%%%%%%%%%%%%%%%%%%%%%%%%%%%%%%%%%%%%%%%%%%%%%%%%%%%%%%
%%%%%%%%%%%%%%%%%%%%%%%%%%%%%%%%%%%%%%%%%%%%%%%%%%%%%%%%%%%%%%%%%%
%%%%%%%%%%%%%%%%%%%%%%%%%%%%%%%%%%%%%%%%%%%%%%%%%%%%%%%%%%%%%%%%%%%
%%%%%%%%%%%%%%%%%%%%%%%%%%%%%%%%%%%%%%%%%%%%%%%%%%%%%%%%%%%%%%%%%%
%%%%%%%%%%%%%%%%%%%%%%%%%%%%%%%%%%%%%%%%%%%%%%%%%%%%%%%%%%%%%%%%%%%
%%%%%%%%%%%%%%%%%%%%%%%%%%%%%%%%%%%%%%%%%%%%%%%%%%%%%%%%%%%%%%%%%%%
%%%%%%%%%%%%%%%%%%%%%%%%%%%%%%%%%%%%%%%%%%%%%%%%%%%%%%%%%%%%%%%%%
%%%%%%%%%%%%%%%%%%%%%%%%%%%%%%%%%%%%%%%%%%%%%%%%%%%%%%%%%%%%%%%%%%%
%%%%%%%%%%%%%%%%%%%%%%%%%%%%%%%%%%%%%%%%%%%%%%%%%%%%%%%%%%%%%%%%%%
%%%%%%%%%%%%%%%%%%%%%%%%%%%%%%%%%%%%%%%%%%%%%%%%%%%%%%%%%%%%%%%%%%
%%%%%%%%%%%%%%%%%%%%%%%%%%%%%%%%%%%%%%%%%%%%%%%%%%%%%%%%%%%%%%%%%%
%%%%%%%%%%%%%%%%%%%%%%%%%%%%%%%%%%%%%%%%%%%%%%%%%%%%%%%%%%%%%%%%%%%
%%%%%%%%%%%%%%%%%%%%%%%%%%%%%%%%%%%%%%%%%%%%%%%%%%%%%%%%%%%%%%%%%%
\begin{figure*}[h!]
\begin{tabular}{c c}
\hspace{-0.3cm}
\includegraphics[width=0.49\textwidth,height=6.6cm]{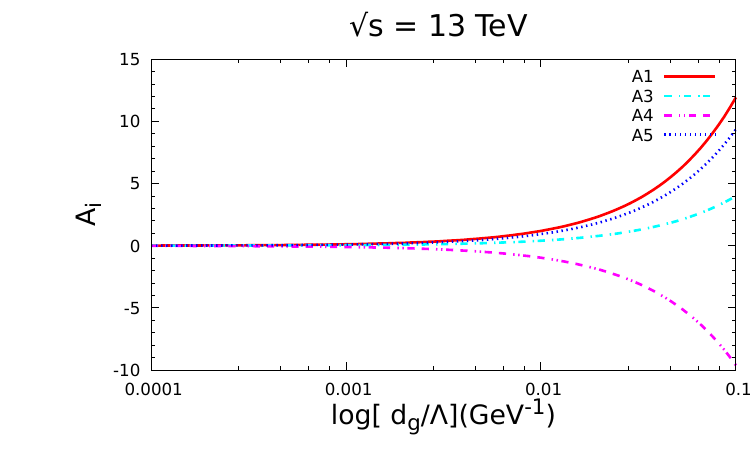}\hspace{-0.3cm}
&\includegraphics[width=0.49\textwidth,height=6.6cm]{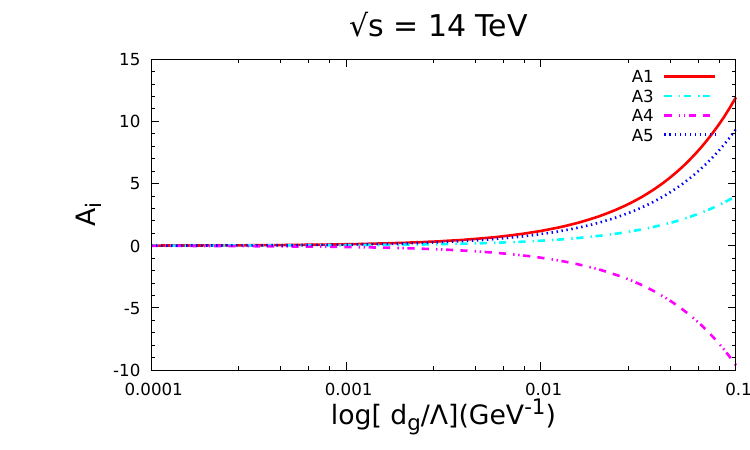}\hspace{-0.3cm}
\end{tabular}
 \caption{Asymmetries vs. $\frac{d_g}{\Lambda}$ for $\sqrt{S}$ = 13 TeV and 14 
TeV energy at LHC.}
\label{Asymmdg/lambda} 
\end{figure*}
%%%%%%%%%%%%%%%%%%%%%%%%%%%%%%%%%%%%%%%%%%%%%%%%%%%%%%%%%%%%%%%%%%%
\\
In Figures \ref{Plot13TeVC1} -- \ref{Plot14TeVC5} the area shown in white is discarded by restricting the contribution in top-pair cross-section to be consistent with the SM within 3$\sigma$ statistical errors whereas the yellow and red regions show possible $d_g - \Lambda$ space allowed at 2.5$\sigma$ and 5$\sigma$ respectively for the given C.M. energy and Luminosities. We have a wide range of $\tilde d_g~(\frac{d_g}{\Lambda})$ values at which we can observe 5$\sigma$ sensitivity at 13 TeV and 14 TeV LHC energies. From the figures, we can get a rough estimate of minimum bound on $d_g$ and $\Lambda$ and can find the lower limit on $\tilde d_g~(\frac{d_g}{\Lambda})$ at 5$\sigma$ C.L..

Finally, we calculate the exact limits on $\tilde d_g$ corresponding to the most promising observable $\mathcal C_1$ at $\sqrt{S}$ = 13 TeV and 14 TeV energy at LHC. The experimental sensitivity at $\sqrt S$ = 13 TeV energy at LHC is found to be 0.2$\%$ at 1$\sigma$ C.L. and the similar value at 5$\sigma$ C.L. would be 1.0$\%$. This translates into the values of $\left|\frac{d_g}{\Lambda}\right|$ of about $ \gtrsim 0.6 \times 10^{-4}$ GeV$^{-1},~0.2 \times 10^{-4}$ at 5$\sigma$ C.L. at 13 TeV C.M. energy with the integrated luminosities of 36.1 fb$^{-1}$, 140 fb$^{-1}$ respectively for observable $\mathcal C_1$. Similarly, at 14 TeV C.M. energy at LHC the value of $\left|\frac{d_g}{\Lambda}\right|$ should be $~\gtrsim 0.6 \times 10^{-5}~\rm{GeV}^{-1},~0.5 \times 10^{-5}~\rm{GeV}^{-1},~0.9 \times 10^{-5}~\rm{GeV}^{-1}$ and $0.1 \times 10^{-4}~\rm{GeV}^{-1}$ at 5$\sigma$ C.L. for the projected luminosities of 0.3 ab$^{-1}$, 1 ab$^{-1}$, 2 ab$^{-1}$ and 3 ab$^{-1}$ respectively. The asymmetries ($\mathcal A_i$) corresponding to observables ($\mathcal C_i$) could also be written as 
%%%%%%%%%%%%%%%%%%%%%%%%%%%%%%%%%%%%%%%%%%%%%%%%%%%%%%%%%%%%%
\begin{eqnarray}
\label{asymmnew}
\mathcal A_{i} &=& \mathcal A_{SM} + b_i \frac{d_g}{\Lambda},
\end{eqnarray}
%%%%%%%%%%%%%%%%%%%%%%%%%%%%%%%%%%%%%%%%%%%%%%%%%%%%%%%%%%%%
where $b_i$ is defined via  
%%%%%%%%%%%%%%%%%%%%%%%%%%%%%%%%%%%%%%%%%%%%%%%%%%%%%%%%%
\begin{eqnarray}
\label{constantb}
 b_i &=& \frac{d\mathcal A_{i}}{d(\frac{d_g}{\Lambda})}. 
\end{eqnarray}    
%%%%%%%%%%%%%%%%%%%%%%%%%%%%%%%%%%%%%%%%%%%%%%%%%%%%%%%%%%%

Figure \ref{Asymmdg/lambda} clearly shows that asymmetries are almost zero up to $10^{-3}$ and then start increasing slowly. It shows that at large $\frac{d_g}{\Lambda}$, sensitivities become quite significant. 
\\
The aim of this article is to set bounds on anomalous CP-violating coupling for a situation when the effects due to such interactions are not visible by just event count, rather these could be probed through the observables considered in our study. We have presented 5$\sigma$ sensitivities for 13 TeV C.M. energy at LHC with the integrated luminosities of 36.1 $\rm fb^{-1}$, 140 fb$^{-1}$ for $\sim$ 19k, 73.5k events per month respectively and predicted that we can achieve 5$\sigma$ sensitivity at 14 TeV LHC energy with projected luminosities of 0.3 ab$^{-1}$, 1 ab$^{-1}$, 2 ab$^{-1}$ and 3 ab$^{-1}$ with $\sim$ 183k, 608k, 1.2M and 1.8M events respectively. The results obtained in our study are based only on statistical uncertainties, systematic uncertainties have not been accounted for. However since it will affect the numerator and denominator in the asymmetry almost equally and therefore it is expected that our results will remain practically unaffected due to the systematic uncertainties. The above finding is also confirmed by earlier studies on such CP asymmetries \cite{deVries:2018mgf,Zhou:1998wz,Lee:2011dqa}. Also in a similar manner, although we had performed our analysis at the leading order, the K-factor due to higher-order QCD corrections will affect the denominators and numerators of all the asymmetries almost equally and will be therefore canceled and hence the asymmetries will remain unchanged against such corrections. It is important to note that our study differs from the earlier studies by taking into account full matrix-element-squared calculation for $pp \to t\bar{t} \to (bl^{+}\nu_l)(\bar{b}l^{-}\bar{\nu_l})$ with $l$ being e and $\mu$. In order to probe the effects of such interactions the earlier studies only considered the leading effects which are linear in nature. Also, we calculated the counting asymmetries in dilepton channel and used $d_g$ and $\Lambda$ as free parameters.

We now compare our results with other relevant works. According to Ref. \cite{Gupta:2009wu}, $5\sigma$ sensitivity of $\left| \tilde d_g \right| < 0.3 \times 10^{-3}$ GeV$^{-1}$ requires 10 fb$^{-1}$ of data at 14 TeV LHC energy. The corresponding estimates are found to be $-0.8\times10^{-4}~<~\tilde d_g~<~0.8\times10^{-4}$ for the top-quark pair production in association with two photons \cite{Etesami:2018mqk} for an integrated luminosity of 3 ab$^{-1}$ and of about $10^{-4}~\rm{GeV}^{-1}$ in the context of $e^-e^+$ collider with a data of about $50$ fb$^{-1}$ \cite{Jezabek:2000gr}. The indirect limits from the EDM measurements are found to be somewhat stringent, e.g. Ref. \cite{Kamenik:2011dk} reports that $\left| \tilde d_g \right| < 1.1 \times 10^{-5} \rm{GeV}^{-1}$ at 90$\%$ C.L. from the measurement of the neutron electric dipole moment.

\section{Summary} \label{summary}

We have analysed the effect of T-odd anomalous couplings of the top-quark with gluons via the top-quark pair production through their semileptonic decay modes at the LHC for $\sqrt{S} = 13$ TeV and 14 TeV using the T-odd observables discussed in Eqs. \ref{oldobs} and \ref{newobs}. These observables are interesting as these do not require full reconstruction of the $t{\bar t}$, rather these require the momenta of the visible final state particles which in our case are $b l^+$ and ${\bar b} l^-$ pairs emerging due to decay of a top and anti-top quarks respectively. The asymmetries corresponding to the T-odd observables have been estimated using Eq. \ref{Asymmetry formula} and are presented in Figures \ref{Plot13TeVC1} -- \ref{Plot14TeVC5} for 13 TeV and 14 TeV LHC energies. Using the largest asymmetry, $\mathcal A_1$ which corresponds to the observable $\mathcal C_1$, we estimated the sensitivity to the CP-violating couplings for $\sqrt{S}$ = 13 TeV energy at LHC with the integrated luminosities of $\int \mathcal L dt = 36.1~\rm{fb}^{-1}$, 140 fb$^{-}$ to be $\bigg|\frac{d_g}{\Lambda}\bigg|~\lesssim~0.29 \times 10^{-4}~\rm{GeV}^{-1}$, 0.52 $\times 10^{-5}~\rm{GeV}^{-1}$ at $3\sigma$ C.L. and $\bigg|\frac{d_g}{\Lambda}\bigg|~\lesssim~0.6 \times 10^{-4}~\rm{GeV}^{-1}$, 0.2 $\times 10^{-4}~\rm{GeV}^{-1}$ at $5\sigma$ C.L. respectively. The corresponding estimates for the HL-LHC with $\sqrt{S}$ = 14 TeV and $\int \mathcal L dt = 0.3~\rm{ab}^{-1}, 1~\rm{ab}^{-1}, 2~\rm{ab}^{-1}$ and $3~\rm{ab}^{-1}$ would yield $\bigg|\frac{d_g}{\Lambda}\bigg|~\lesssim~0.39 \times 10^{-5}~\rm{GeV}^{-1},~0.11 \times 10^{-4}~\rm{GeV}^{-1},~0.13 \times 10^{-4}~\rm{GeV}^{-1}$ and $0.14 \times 10^{-4}~\rm{GeV}^{-1}$ at $3\sigma$ C.L. and $\bigg|\frac{d_g}{\Lambda}\bigg|~\lesssim~0.6 \times 10^{-5}~\rm{GeV}^{-1},~0.5 \times 10^{-5}~\rm{GeV}^{-1},~0.9 \times 10^{-5}~\rm{GeV}^{-1}$ and $0.1 \times 10^{-4}~\rm{GeV}^{-1}$ at $5\sigma$ C.L. respectively. These results have been summarised in Table \ref{sens_results} and seem to be setting stringent bounds on the CP-violating couplings of the top-quark and therefore a detailed experimental investigation is worthwhile to shed light on such CP-violating couplings of the top-quark.

%%%%%%%%%%%%%%%%%%%%%%%%%%%%%%%%%%%%%%%%%%%%%%%%%%%%%%%%%%%%
\begin{table}[h!]
\centering
\scalebox{1.2}{
\renewcommand{\arraystretch}{1.4}
\begin{tabular} {|c|c|c|c|}
\hline
$\sqrt{S}~(\rm{TeV})$ & $\int \mathcal L dt$ & \multicolumn{2}{|c|}{$\left|\frac{d_g}{\Lambda}\right|~(\rm{in~GeV}^{-1})$}\\
\hline
 &  & $\rm{at}~3\sigma~\rm{C.L.}$ & $\rm{at}~5\sigma~\rm{C.L.}$\\
  \hline
13 & 36.1 fb$^{-1}$ & 0.29 $\times 10^{-4}$ & 0.6 $\times 10^{-4}$\\
 & 140 fb$^{-1}$ & 0.52 $\times 10^{-5}$ & 0.2 $\times 10^{-4}$\\                                
\hline
14 (HL-LHC) & 0.3 ab$^{-1}$ & 0.39 $\times 10^{-5}$ & 0.6 $\times 10^{-5}$\\
 & 1.0 ab$^{-1}$ & 0.11 $\times 10^{-4}$ & 0.5 $\times 10^{-5}$\\
 & 2.0 ab$^{-1}$ & 0.13 $\times 10^{-4}$ & 0.9 $\times 10^{-5}$\\
 & 3.0 ab$^{-1}$ & 0.14 $\times 10^{-4}$ & 0.1 $\times 10^{-4}$\\
 \hline
\end{tabular}}
\caption{Sensitivity to CP-violating anomalous couplings at 3$\sigma$ C.L. and 5$\sigma$ C.L. in the process $pp \to t\bar{t} \to (bl^{+}\nu_l)(\bar{b}l^{-}\bar{\nu_l})$ at $\sqrt{S}$ = 13 TeV energy at LHC with the integrated luminosities of 36.1 fb$^{-1}$, 140 fb$^{-1}$ and HL-LHC with $\sqrt{S}$ = 14 TeV energy with the projected luminosities of 0.3 ab$^{-1}$, 1 ab$^{-1}$, 2 ab$^{-1}$ and 3 ab$^{-1}$.}
\label{sens_results}
\end{table}
%%%%%%%%%%%%%%%%%%%%%%%%%%%%%%%%%%%%%%%%%%%%%%%%%%%%%%%%%%%%%%

\begin{acknowledgements}
This work was supported in part by University Grant Commission under a 
Start-Up Grant no. F30-377/2017 (BSR). We thank Ravindra Yadav for his 
assistance regarding high-performance computing, Manjari Sharma and 
Surabhi Gupta for some valuable discussions. We acknowledge the use of 
cluster computing facility at the ReCAPP, HRI, Allahabad, India during 
the initial phase of the work.
\end{acknowledgements}

\end{document}